\documentclass[letterpaper,twocolumn,showpacs,prb,superscriptaddress,email]{revtex4-1}

\usepackage[pdftex]{graphicx}  
\usepackage{amssymb}
\usepackage{amsmath,amsfonts,latexsym}
\usepackage{array,tabularx,color}
\usepackage{dcolumn}               
\usepackage[normalem]{ulem}
\usepackage{comment}
\usepackage{bm}
\usepackage[latin1]{inputenc}
\usepackage{color}

\begin{document}
\title{Interaction-Shaped Vortex-Antivortex Lattices in Polariton Fluids}

\author{R. Hivet} \affiliation{Laboratoire Kastler Brossel, Universit\'e Pierre et Marie Curie, Ecole Normale Sup\'erieure et CNRS, UPMC case 74, 4 place Jussieu, 75005 Paris, France}

\author{E. Cancellieri} \affiliation{Laboratoire Kastler Brossel, Universit\'e Pierre et Marie Curie, Ecole Normale Sup\'erieure et CNRS, UPMC case 74, 4 place Jussieu, 75005 Paris, France}

\author{T. Boulier} \affiliation{Laboratoire Kastler Brossel, Universit\'e Pierre et Marie Curie, Ecole Normale Sup\'erieure et CNRS, UPMC case 74, 4 place Jussieu, 75005 Paris, France}

\author{D. Ballarini} \affiliation{CNB@UniLe, Istituto Italiano di Tecnologia, via Barsanti, 73100 Arnesano (Lecce), Italy}

\author{D. Sanvitto} \affiliation{CNB@UniLe, Istituto Italiano di Tecnologia, via Barsanti, 73100 Arnesano (Lecce), Italy}

\author{F. M. Marchetti} \affiliation{Departamento de Fisica Teorica de la Materia Condensada y Condensed Matter Physics Center (IFIMAC), Universidad Autonoma de Madrid, Spain}

\author{M. H. Szymanska} \affiliation{Department of Physics and Astronomy, University College London, Grower Street, London WC1E6BT, United Kingdom}

\author{C. Ciuti} \affiliation{Laboratoire Mat\'eriaux et Ph\'enom\'enes Quantiques, UMR 7162, Universit\'e Paris Diderot-Paris 7 et CNRS, 75013 Paris, France}

\author{E. Giacobino} \affiliation{Laboratoire Kastler Brossel, Universit\'e Pierre et Marie Curie, Ecole Normale Sup\'erieure et CNRS, UPMC case 74, 4 place Jussieu, 75005 Paris, France}

\author{A. Bramati} \affiliation{Laboratoire Kastler Brossel, Universit\'e Pierre et Marie Curie, Ecole Normale Sup\'erieure et CNRS, UPMC case 74, 4 place Jussieu, 75005 Paris, France}

\begin{abstract}
Topological defects such as quantized vortices are one of the most striking manifestations of the superfluid nature of Bose-Einstein condensates and typical examples of quantum mechanical phenomena on a macroscopic scale. Here we demonstrate the formation of a lattice of vortex-antivortex pairs and study, for the first time, its properties in the non-linear regime at high polarion-density where polariton-polariton interactions dominate the behaviour of the system. In this work first we demonstrate that the array of vortex-antivortex pairs can be generated in a controllable way in terms of size of the array and in terms of size and shape of it fundamental unit cell. Then we demonstrate that polariton-polariton repulsion can strongly deform the lattice unit cell and determine the pattern distribution of the vortex-antivortex pairs, reaching a completely new behaviour with respect to geometrically generated vortex lattices whose shape is determined only by the geometry of the system.
\end{abstract}

\maketitle

\section{Introduction}
Quantized vortices are topological excitations characterized by the vanishing of the field density at a given point, the vortex core, and by the quantized winding of the field phase from 0 to $2\pi m$ around it (with m an integer number). Together with solitons they have been extensively studied \cite{onsager_statistical_1949,feynman_vol._1955,nye_dislocations_1974} and observed in non-linear optical systems \cite{desyatnikov_optical_2005}, superconductors \cite{essmann_direct_1967}, superfluid $^4$He \cite{yarmchuk_observation_1979}, vertical-cavity surface-emitting lasers \cite{scheuer_optical_1999}, and more recently in cold atoms \cite{madison_vortex_2000,denschlag_generating_2000,khaykovich_formation_2002} where, as predicted by Abrikosov \cite{abrikosov_magnetic_1957}, vortices tend to arrange in triangular lattices due to their mutual interactions. Finally, in recent years, the study of vortices and vortex lattices has attracted much attention also in the field of coherent light-matter systems.

Semiconductor microcavities can be designed to strongly couple cavity photons to quantum well excitons. The eigenstates of this system are called exciton-polaritons and are characterised by specific properties such as low effective mass, inherited from their photonic component, and strong non-linear interactions due to their excitonic part. Moreover, polaritonic systems are easily controllable by optical techniques and, due to their finite lifetime are ideal systems to study out of equilibrium phenomena \cite{deng_exciton-polariton_2010,carusotto_quantum_2013}. In analogy with the atomic case \cite{dalfovo_theory_1999,burger_superfluid_2001} the superfluid behaviour of polaritonic Bose-Einstein condensates \cite{kasprzak_boseeinstein_2006} has been of great theoretical interest \cite{carusotto_probing_2004,wouters_superfluidity_2010,cancellieri_superflow_2010,cancellieri_frictionless_2012} and has been experimentally confirmed by the suppression of scattering in the case of a polariton fluid flowing past a defect \cite{amo_collective_2009,amo_superfluidity_2009} and by the persistence of circular quantized currents \cite{sanvitto_persistent_2010}. In particular, cavity-polariton systems have been predicted and shown to undergo formation of stable vortices \cite{lagoudakis_quantized_2008} and half-vortices \cite{lagoudakis_observation_2009,flayac_topological_2010} as well as formation of single vortex-antivortex (V-AV) pairs \cite{roumpos_single_2011,nardin_hydrodynamic_2011,tosi_onset_2011}. More recently, Amo et al. \cite{amo_polariton_2011} and Hivet et al. \cite{hivet_half-solitons_2012} have demonstrated the nucleation of hydrodynamic solitons and half-solitons in resonantly pumped polaritons flowing against an extended obstacle. The formation of lattices of vortex and of vortex-antivortex pairs has been theoretically predicted and experimentally studied for cavity-polaritons \cite{gorbach_vortex_2010,keeling_spontaneous_2008,liew_generation_2008}, and its appearance has been observed in the case of patterns induced by metallic deposition on the surface of the cavity \cite{kusudo_stochastic_2012}, and in the case in which the interplay between the excitation shape and the underlying disordered potential is able to pin the position of the vortices allowing their detection in time-integrated experiments \cite{manni_2013}.

In the present work, we use a continuous wave laser to resonantly inject polaritons outside of a masked region \cite{pigeon_hydrodynamic_2011} and to observe vortex lattices trapped by an optically controllable potential that, at the same time, stimulates the lattice formation. Here, the resonant pumping configuration allows for a fine tuning of the polariton density but does not generate an excitonic reservoir. For this reason, we can address theoretically and observe experimentally, for the first time, the effects of the polariton-polariton non-linear interactions on the shape of the lattice of vortices. This is in contrast with what was possible to observe with an out-of-resonance setup \cite{tosi_geometrically_2012,cristofolini_optical_2013}. In these latter cases, either the shape of the lattice was completely determined by the geometry of the pumping scheme \cite{tosi_geometrically_2012}, or a transition to trapped states was observed by bringing the pump spots closer to each other \cite{cristofolini_optical_2013}. In these experiments the excitonic reservoir plays a fundamental role by determining the characteristic length and the shape of the formed vortex arrays, and in generating the potential where the polariton condensate is trapped, leading to the disappearance of the vortex lattice.

The paper is organized in three main sections, in the first section we describe the setup used to perform the experiments, in the second we highlight our main results distinguishing between two main regimes: the linear regime at low polariton densities, and the non-linear regime at high polariton densities. In this section we address ways to control the shape and the size of the generated array and show that a new regime can be reached in which polariton-polariton interactions determine the shape of the array. Finally, in the last section, we derive our conclusions.

\section{Experimental Setup}
In our experiment we use a high finesse GaAs microcavity ($F=3000$) with a polariton lifetime $\tau=15$ ps and a Rabi splitting of 5.1 meV \cite{houdre_coherence_2000,adrados_motion_2011,sanvitto_all-optical_2011} (see Appendix A: {\it Sample description}). The microcavity is excited with a continuous-wave single-mode Ti:Sa laser quasi resonant with the lower polariton branch at 837 nm. The pump laser is circularly polarized in order to avoid any effect due to spin-dependent interactions \cite{vladimirova_polariton-polariton_2010}. The output beam can be made to interfere with a reference beam of constant phase from the same laser before being collected on a CCD camera, thus allowing for the reconstruction of the phase of the fluid with simple numerical treatment (see Appendix B: {\it Phase extraction procedure}).
The experimental observations have been performed at 10 K in transmission configuration and for different positions of the laser spot on the cavity that correspond to different exciton-photon detunings $\delta_{exc-photon}=\omega_{X}(k=0)-\omega_C(k=0)$, where $\omega_{X(C)}(k=0)$ are the excitonic and photonic energies at normal incidence ($k=0$, where $k$ is the projection of the light wavevector on the plane of the microcavity). In order to observe vortices, it is critical to let the polariton phase evolve freely after they have been injected by the laser. In previous studies polaritons were observed in a region of the microcavity where they have moved away from the laser spot, which was efficiently limited by a mask. This technique, ensuring a free evolution of the phase, allowed for the observation of solitons \cite{amo_polariton_2011}, half-solitons \cite{hivet_half-solitons_2012} and vortices \cite{pigeon_hydrodynamic_2011}.

\begin{figure}
\includegraphics[width=8cm]{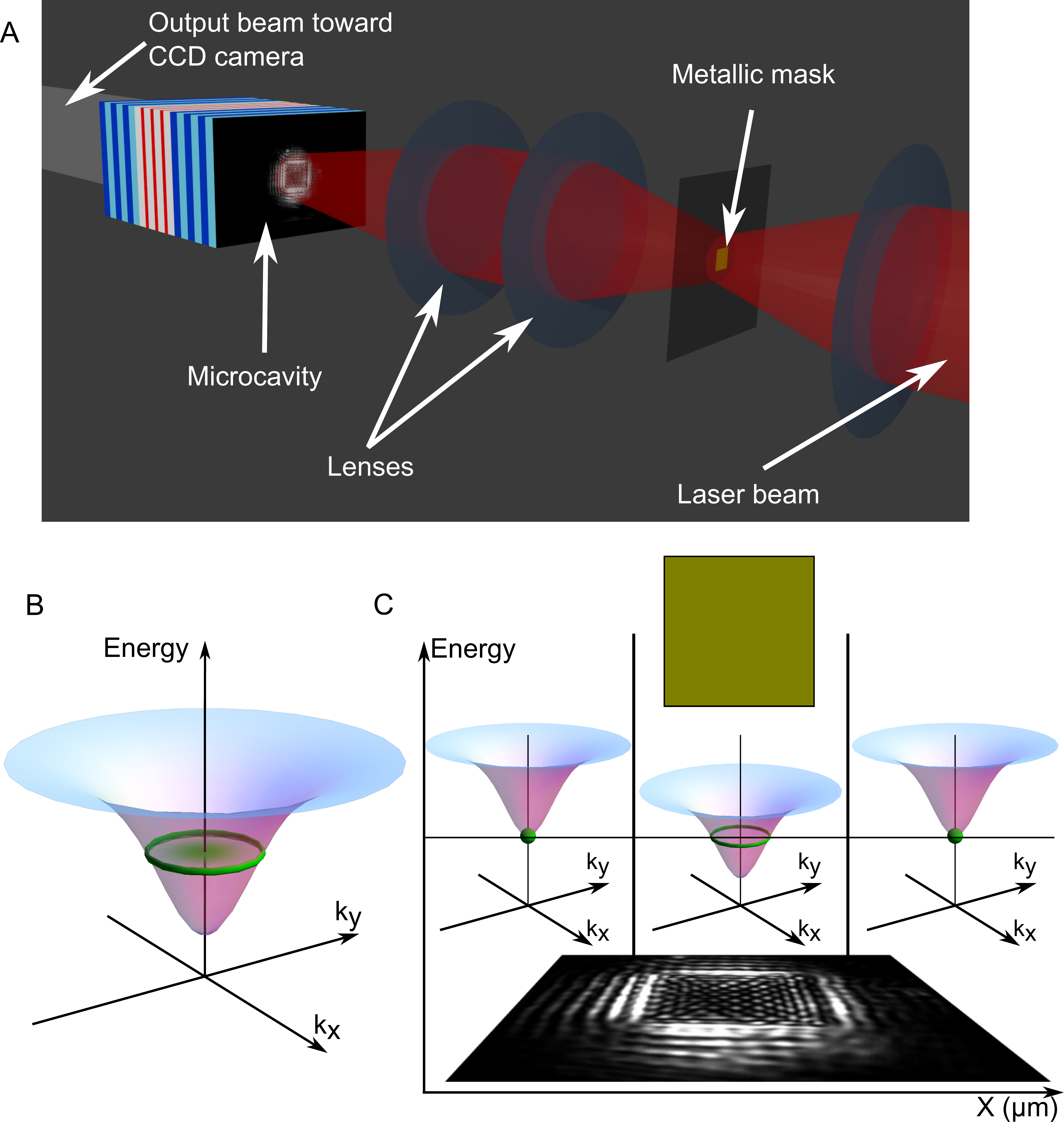}
\caption{\label{fig1}A: Illustration of the excitation scheme. The waist of the beam is first imaged on a 0.5 mm square metallic mask which is then imaged at $k=0$ on the microcavity surface. The image of the mask generates a 45 $\mu$m side square region at the centre of bright Gaussian spot of 110 $\mu$m of full width at half maximum. B: At low polariton density, the laser pump energy (green spot) is blue-shifted with respect to the bare LP at k=0 and crosses the LP at a k corresponding to the Rayleigh ring. Therefore, polaritons diffracted by the mask borders can be injected into the cavity at a k corresponding to $k_R$ of the Rayleigh ring (green ring). C: At high laser intensities, the lower polariton branch outside of the masked region (left and right) is renormalized due to polariton-polariton interactions and therefore polaritons are injected at $k=0$ (green dot). Inside the masked region (central region), the polariton density being lower, the polariton branch is not renormalized and therefore, when polaritons enter by diffusion acquire a momentum (green ring) in order to conserve the energy.}
\end{figure}

We focus the laser beam on a metallic gold-coated mask smaller than the beam waist that locally sets the laser intensity to zero. Using masks with different shapes and sizes (such as triangles and squares), it is possible to obtain laser beams with zero intensity regions of different shapes. This partially obscured waist is then imaged on the microcavity therefore creating a polariton fluid outside a dark region whose size and shape can be set at will (fig. \ref{fig1}A). In this experiment the beam waist has a diameter of 110 $\mu$m with a square dark region in the centre with a side of 45 $\mu$m, and an energy set above the bare polariton energy at $k=0$ for the considered point on the microcavity. Therefore the laser energy crosses the bare lower polariton dispersion curve at a non-zero value of $|k|=|k_R|$, (fig. \ref{fig1}B). This crossing has the shape of a ring, corresponding to the Rayleigh ring. Since the beam is set at normal incidence this value of $k$ is not present in the laser beam. However, due to the presence of the mask, photons are diffracted with various values of $|k|$ perpendicular to the edges and can enter the cavity when they have the right $|k|= |k_R|$ value. Polaritons then enter into the dark region flowing with four directions perpendicular to the mask edges. When the laser intensity is increased, in the bright region of the spot, the lower polariton branch (LPB) is blue-shifted into resonance with the laser energy at $k=0$ due to the high polariton density (left and right regions of fig. \ref{fig1}C). In the dark region of the spot (central region of fig. \ref{fig1}C) the LPB is not renormalized because the polariton density is low. Therefore, polaritons created at the edge of the bright region can travel into the dark region getting a momentum $|k|=|k_R|$ that conserves their energy \cite{wertz_spontaneous_2009}. As in the non resonant excitation case, polaritons flow into the dark region with four directions perpendicular to the mask edges.

The system being completely symmetric, it is bound to keep a total angular momentum equal to zero and, therefore, the number of generated vortices is bound to be always equal to the number of antivortices. In this sense vortices and antivortices are always generated in pairs although they do not necessarily form bound states. Here vortices and antivortices are characterized by a $2\pi$ or $-2\pi$ rotation of the phase of the wavefunction around the core of the topological defect where the polariton density goes to zero (See inset of fig. \ref{fig2}B).

\section{Results}
In order to better highlight the mechanism lying beneath the formation of the V-AV lattice and the role of polariton-polariton interactions, we study the system as a function of the polariton density. We identify two different regimes: a linear regime at low polariotn density; and a non-linear regime at high polariton density. The linear regime is characterised by the polariton density lying on the lower branch of the bistability curve \cite{baas_optical_2004} everywhere in space, and its behaviour is completely linear. The non-linear regime corresponds to a polariton density that is on the upper branch of the bistability curve outside the masked region and that gradually decreases until it reaches the lower branch at the center of the trap. In this regime polariton-polariton interactions dominate the behaviour of the system.

\subsection{Linear regime.}
\begin{figure}
\includegraphics[width=8cm]{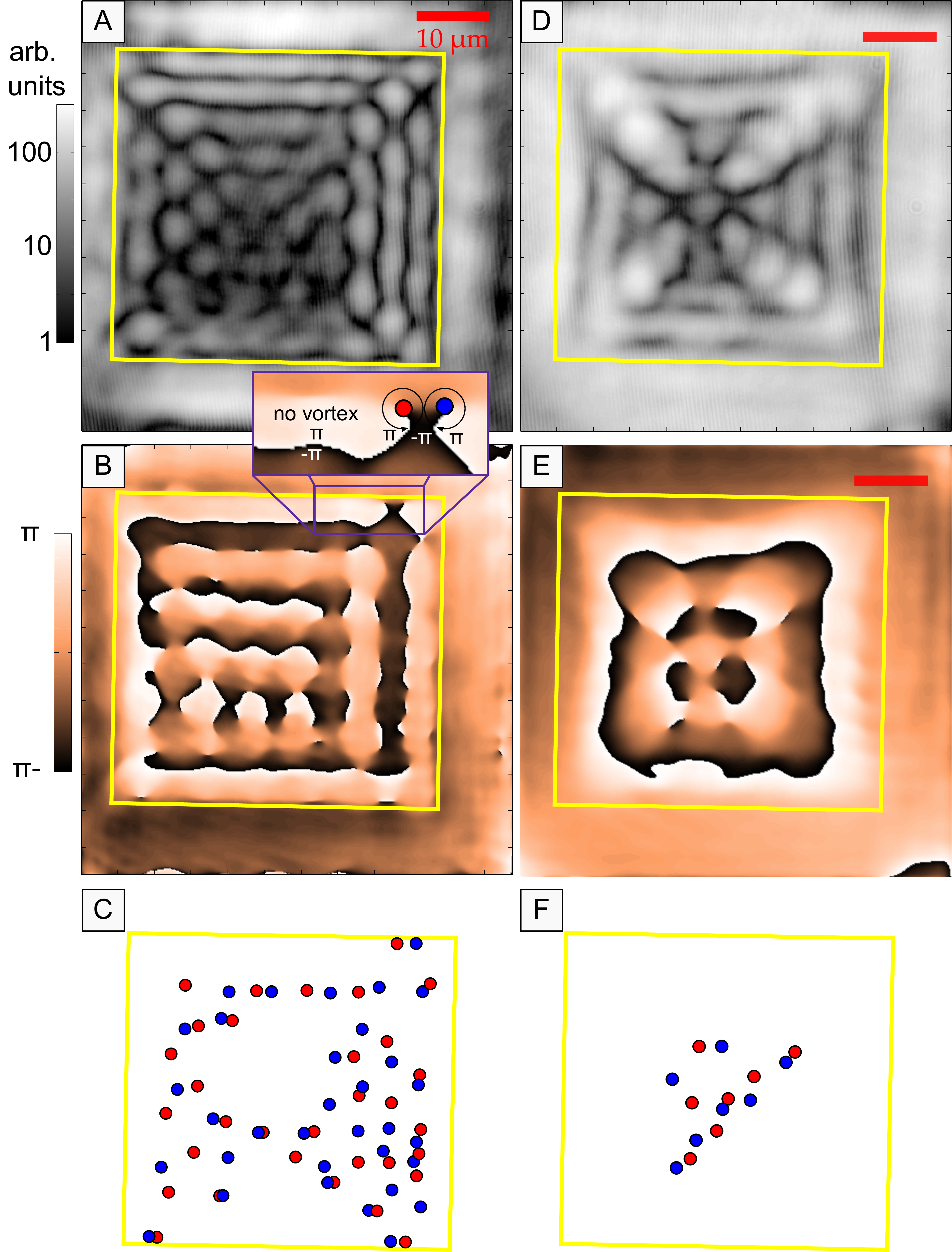}
\caption{\label{fig2} Experimental real space emission intensity, phase and vortex distribution in the linear and non-linear regimes. The microcavity has a detuning $\delta_{exc-photon}=0.32$ meV and the pump laser has a blue shift from the LPB at $k=0$ of $\delta_{pump-LPB}=0.6$ meV. A-C: Linear regime, pump power 1 mW. In this regime the injected wavenumber is determined by the radius of the Rayleigh ring ($k_R=1.1\ \mu$m$^{-1}$). A, B, and C correspond respectively to the experimental real space emission, its corresponding phase and the set of vortices (red dots) and anti-vortices (blue dots) formed inside the trapped region (in yellow). The positions of the vortices are obtained from a detailed analysis of the phase map B. The inset shows a vortex (in red) with phase varying counter-clockwise from $-\pi$ to $\pi$ around the core, and an antivortex (in blue). The wavenumber characterizing the size of the array is about $k_L=0.9\ \mu$m$^{-1}$. The difference between $k_R$ and $k_L$ is attributed to uncertainty in the measurement. D-F: Non-linear regime with pump power equal to 35 mW. A strong modification of the vortex pattern is observed along with the disappearance of V-AV pairs from the borders. The greyscale used for real space emission is logarithmic while the colour scale for phase diagrams is linear from $-\pi$ to $\pi$.}
\end{figure}

Figure \ref{fig2}A shows the experimental real space distribution of the light transmitted by the cavity when a square mask partially blocks the pumping beam with pump power of 1 mW. The transmitted beam is made to interfere with a reference beam and the phase of the interferogram (fig. \ref{fig2}B) is analysed in details in order to identify the position of the vortices (fig. \ref{fig2}C). Since in this case the laser intensity and the polariton density are low, polariton-polariton interactions play a negligible role and the system is analogous to the case of overlapping laser beams interfering in a medium with linear dispersion. In this first case, where a square mask is used, polaritons mainly flow from the four sides of the mask towards the centre and the polaritonic system corresponds to the case of four plane waves coming from four orthogonal directions. These flows generate an interference pattern with a clearly identifiable wave vector $k_L$ that, from now on, we define as the vortex-lattice characteristic wavevector. However, while for the case of four plane waves only a square interference pattern without any vorticity would appear, here V-AV pairs can form thanks to the non uniform density distribution. The finite lifetime of cavity polaritons, together with the fact that the local polariton density and the local polariton velocity are determined by the sum of the polariton flows coming from the different sides of the mask, induces a non-homogeneous polariton distribution that, in turns, changes the direction of the polariton flows and allows the formation of vortices and antivortices.

Note that while a regular squared interference pattern generated by four plane waves is clearly visible in the experiment (figures \ref{fig2}A and \ref{fig2}B) in the entire masked region, figure \ref{fig2}C shows that there are areas (center and bottom) where vortices and antivortices do not appear. This is mainly due to the unavoidable presence of disorder and defects in the microcavity sample that might inhibit the formation of V-AV pairs; on the other hand if the core of a vortex is very close to the core of an antivortex the experimental resolution in the phase measurements might be insufficient to resolve them. The comparison of these results with the Gross-Pitaevskii based simulations in figure \ref{fig3} (for more details on the theoretical model see Appendix A) shows that a regular interference pattern, with a characteristic wave vector $k_L=0.9$ $\mu m^{-1}$ similar to the one observed in figure \ref{fig2}, is clearly observed also in the numerical simulations, both in the emission intensity map (figure \ref{fig3}A) and in the phase map (figure \ref{fig3}B). Despite these similarities one can see that the actual vortex distribution in figure \ref{fig3}C slightly differs from the one of figure \ref{fig2}C, this is due to the aforementioned imperfections of the real experimental system that can not be entirely reproduced by the simulations.

\subsubsection{Control of the size of the unit cell}
\begin{figure}
\includegraphics[width=8cm]{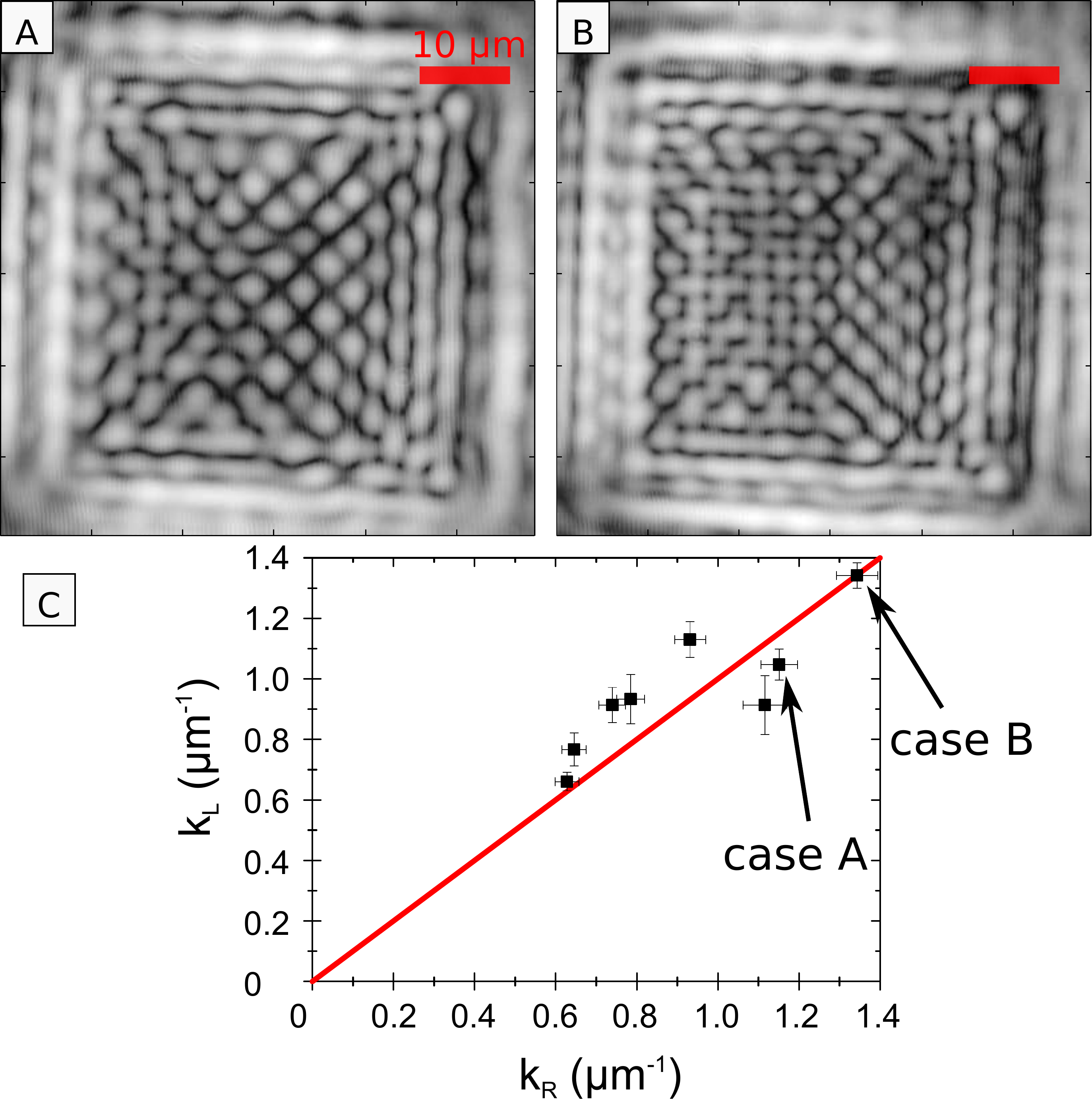}
\caption{\label{fig1s} A-B: Real space emission in the case of a $45\ \mu$m square mask in the linear regime with $\delta_{exc-photon}=0.32$ meV and using different laser wavelengths. The wavelengths used are $\lambda_A=836.40$ nm and $\lambda_B=836.36$ nm respectively inducing a lattice width unit cell of $6\ \mu$m and $4.7\ \mu$m. The lattices size correspond to momenta $k^A_L=1.0\ \mu$m$^{-1}$ and $k^B_L=1.34\ \mu$m$^{-1}$ while the corresponding momenta from the Rayleigh rings are $k^A_R=1.15\ \mu$m$^{-1}$ and $k^B_R=1.34\ \mu$m$^{-1}$. In C we compare, for several different cases, the measured $k_L$ with the corresponding measured $k_R$. Clearly, apart of the differences due to uncertainty in the experimental evaluation, the two values are equal.}
\end{figure}

As the $|k_L|$ of the polaritons flowing into the trap is determined by the detuning between the energy of the pump and the bare LP at normal incidence, a change of the laser frequency allows for a fine tuning of $|k_L|$ and therefore for the control of the lattice unit cell size: an increase (decrease) of the pump frequency corresponds to an increase (decrease) of the momentum of the injected polaritons and therefore to smaller (larger) size of the unit cell. Figures \ref{fig1s}-A and \ref{fig1s}-B represent the transmitted light in the same region of the microcavity in the case where only the frequency of the laser has been changed. The increase (from A to B) of the laser frequency results in the increase of the momentum of the injected polaritons and, therefore, in the decrease of the size of the unit-cell. To study the relation between the unit cell size and the laser-LP detuning, we extract, for several detunings, the characteristic wave vector of the lattice $k_L$ and compare it with the $k$-vector of the corresponding Rayleigh ring $k_R$ given by the relation $E_{LPB}(k_R)=E_{laser}$. The results, shown in Figure \ref{fig1s}-C, demonstrate a linear dependence between the two $k$-vectors.

\subsubsection{Control of the shape of the unit cell}
\begin{figure}
\includegraphics[width=8cm]{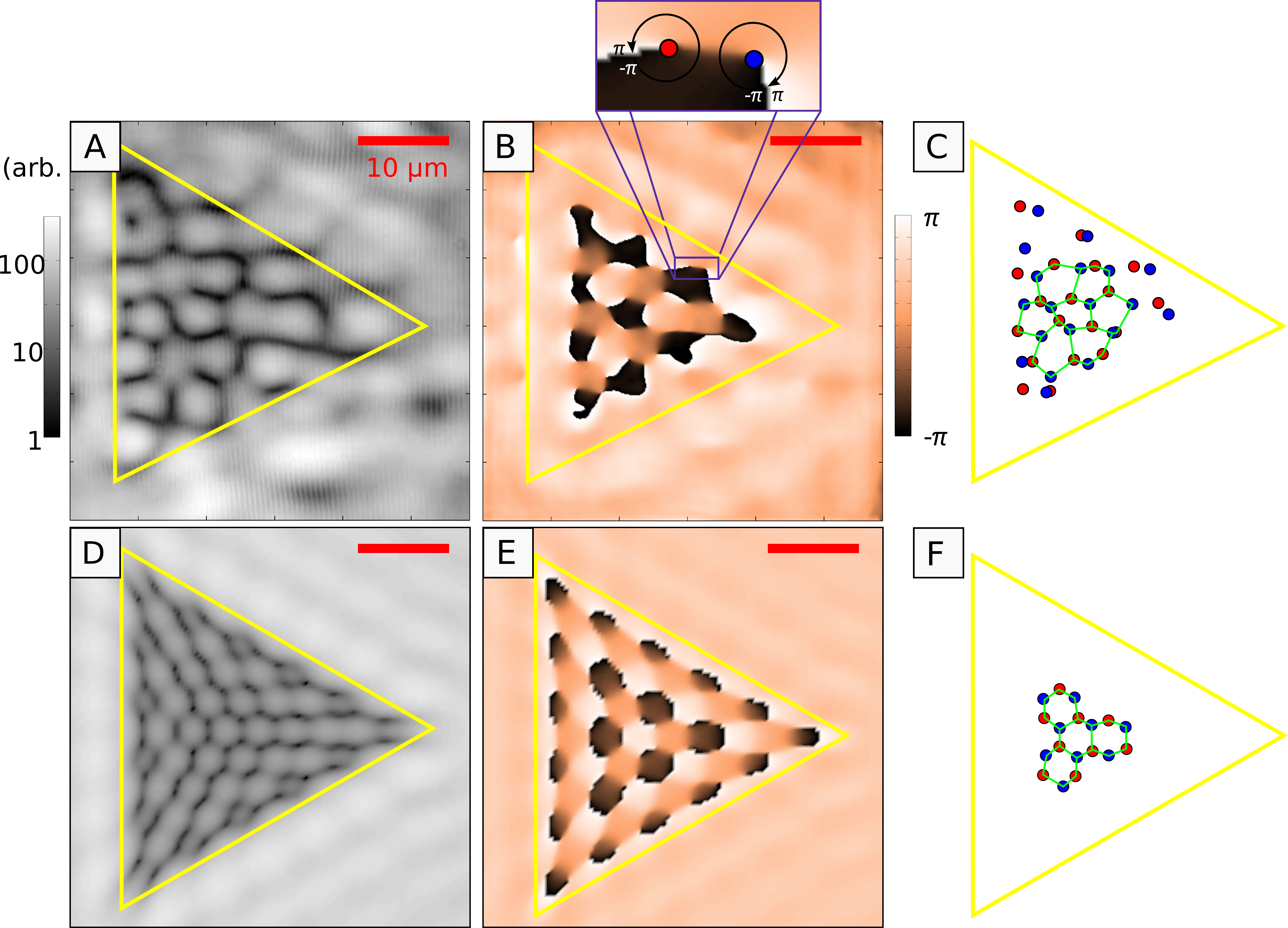}
\caption{\label{fig2s} A-B: Experimental real space emission output for a $45\ \mu$m side triangular mask and its corresponding phase. The yellow triangle represents the mask boundaries. In figure B the inset shows the phase rotating clockwise from $-\pi$ to $\pi$ around the core of an anti-vortex (in blue), together with the clockqise rotating phase of a vortex (in red). C: Representation of the hexagonal lattice of vortices (in red) and anti-vortices (in blue) corresponding to a wavevector $k_L=0.75\ \mu$m$^{-1}$. D-F: Theoretical simulations performed in the same conditions of excitations. The data has been taken at $\delta_{exc-photon}=1.68$ meV, and the pump laser has a blue shift from the LPB at $k=0$ of $\delta_{pump-LPB}=0.16$ meV corresponding to an injected momentum of $k_R=0.65$ $\mu$m$^{-1}$.}
\end{figure}

While the $|k|$ of the injected polaritons is governed by the detuning between the laser and the bare LP branch, the direction of the polaritons is determined by the shape of the mask used to block part of the laser spot. Therefore the shape of the unit cell can be controlled by changing the shape of the maks. In the case of a triangular mask (Figure \ref{fig2s}), when the laser intensity and the polariton density are low and polariton-polariton interactions play a negligible role, the polaritonic fluid forms an array of vortex-antivortex pairs arranged in a hexagonal unit cell as expected for the superposition of three laser beams \cite{masajada_optical_2001}. Figure \ref{fig2s} compares the experimental and theoretical output for a $35\ \mu$m side almost equilateral triangle and its corresponding phase. In the experimental output we can recognize up to 8-9 unit cells in very good agreement with the Gross-Pitaevskii based simulations. In this regime the generation of the lattice is driven by the interferences between polaritons coming from each side of the triangle and flowing toward the centre of the trap. The result is therefore analogous to the case of three overlapping laser beams interfering in a linear medium. Finally, let us mention that changing the size of the mask for fixed polariton wavevector $|k|$ and fixed mask shape results in a change of the overall lattice size only, while size and shape of the unit cells remain unchanged.

\subsection{Non-linear regime.}
\begin{figure}
\includegraphics[width=8cm]{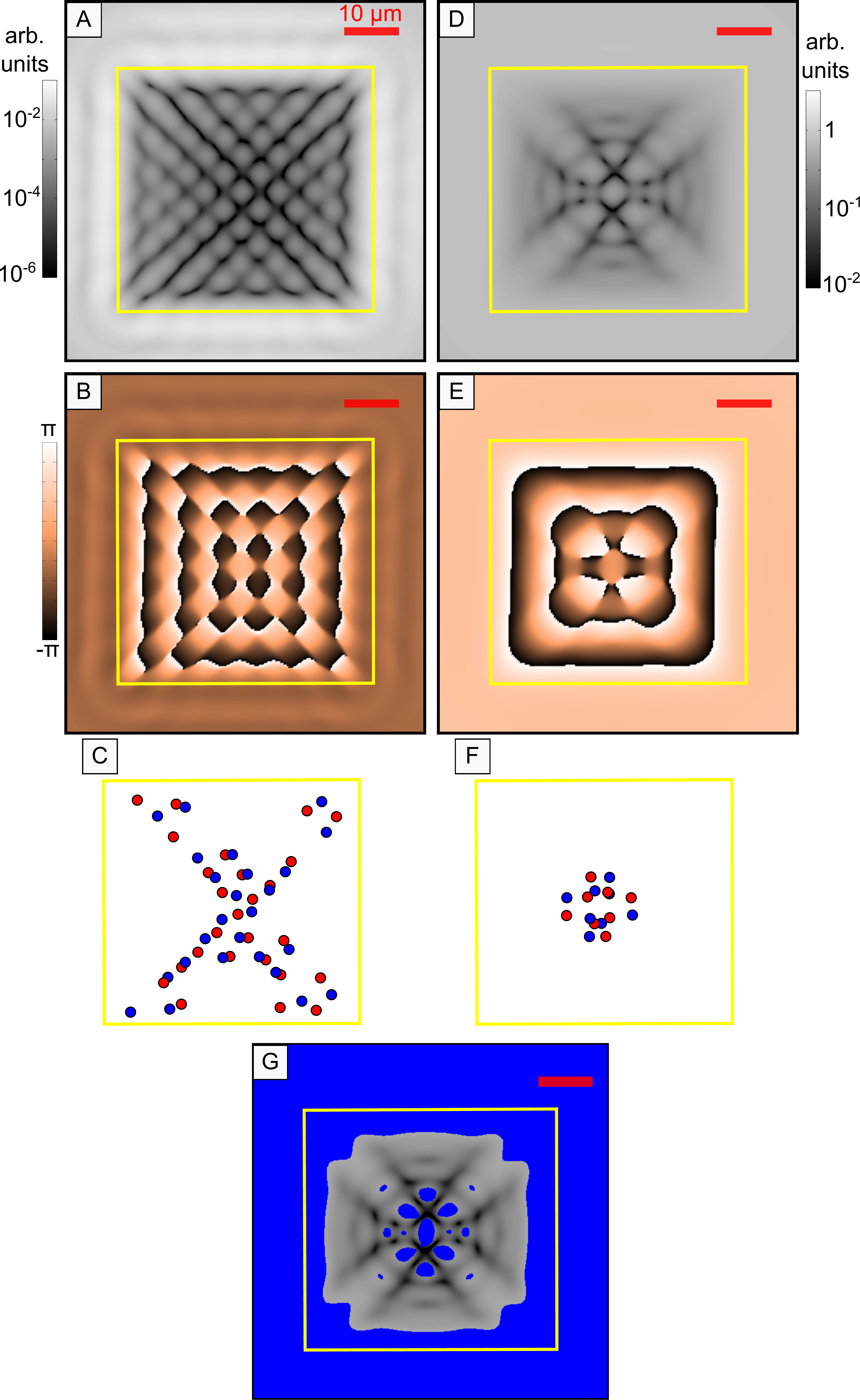}
\caption{\label{fig3} Numerical real space images in the same conditions as in fig. 2. Two sides of the square mask are set to be not perfectly parallel in order to simulate unavoidable asymmetries of the experimental setup. A-C: Linear regime. D-G: High density regime. G: Mach-number chart showing in blue the regions where the Mach number is smaller than 1 (i.e. the fluid is in a subsonic condition) while the real space emission D is reproduced in grey in regions where the Mach number is larger than 1.}
\end{figure}

Figures 2D-F show the experimental polariton distribution in the case of high laser power (pump power has been increased from 1 mW to 35 mW) for the same mask, same position in the cavity and same laser frequency as in fig. \ref{fig2}A-C. Comparing the polariton distribution at low density (fig. \ref{fig2}A-C) with the one at higher density (fig. \ref{fig2}D-F) we see a drastic change in the distribution of the topological defects. At high densities, the repulsion between polaritons not only leads to an enlargement and to a deformation of the lattice unit cell, which is mostly due to a blue shift of the polariton energy, but also leads to a drastic change in the vortex distribution. In particular, as discussed below, the disappearence of the vortices from the outer part of the trap is linked to quantum fluid properties of the polariton condensate. This behaviour is remarkably consistent with the theoretical simulations shown in fig. \ref{fig3}D-F for the same set of parameters as in the experimental case.

To better understand the vortex distribution, we study the correlations between the disappearance of the array of V-AV pairs and the subsonic character of the fluid. Since the distribution of the polariton fluid is not homogeneous, one has to define a local speed of sound $c_s({\bf r})=\sqrt{\hbar g_{LP} |\Psi_{LP}({\bf r})|^2/m_{LP}}$ where $|\Psi_{LP}({\bf r})|^2$, $g_{LP}$ and $m_{LP}$ are the local polariton density, the coupling constant and the polariton mass. In the local density approximation \cite{wouters_spatial_2008}, $c_s({\bf r})$ corresponds to the speed of sound defined in the case of high densities (i.e. when the polariton density lies on the upper branch of the bistability curve and the Bogoliubov dispersion of the collective excitations is linear \cite{ciuti_quantum_2005}) and therefore hereinafter we will simply call it {\it generalized speed of sound}. To study the existence of these correlations we evaluate the Mach number ($M({\bf r})$) defined as the ratio between the speed of the fluid  $v_f({\bf r})=\hbar |{\bf k}({\bf r})|/m_{LP}$ (locally evaluated as the derivative of the phase at the point $({\bf r})$, and the generalized speed of sound ($c_s({\bf r})$):
\begin{equation}
\label{eqM}
M({\bf r})=\frac{v_f({\bf r})}{c_s({\bf r})}=\frac{\hbar |{\bf k}({\bf r})|/m_{LP}}{\sqrt{\hbar g_{LP} |\Psi_{LP}({\bf r})|^2/m_{LP}}}.
\end{equation}

Figure \ref{fig3}G shows the Mach-number chart corresponding to the case of fig. \ref{fig3}D-F: in the region inside the trap but close to the borders, where V-AV pairs have disappeared, the system is in a subsonic regime due to its high density. Since a subsonic fluid cannot sustain strong phase modulations \cite{landau_chapter_1986}, V-AV pairs merge in the regions where the fluid becomes locally subsonic. Note that the disappearance of the V-AV pairs can not be ascribed to a simple renormalization of the lower polariton branch since this mechanism would lead to an increase of the lattice characteristic length. This is not the case in our system since the periodicity related to this characteristic length is still visible in Figures \ref{fig2}D and \ref{fig3}D. In our case V-AV pairs disappear because the increase of the polariton density forces vortices to overlap with antivortices and therefore to annihilate. On the other hand, as polaritons move towards the centre of the trap, due to their finite lifetime, their density decreases and so does the sound velocity of the fluid so that the fluid becomes mainly supersonic ($M({\bf r}) > 1$). In this inner region, V-AV pairs can survive and the array is only slightly deformed. Therefore, the finite lifetime of cavity-polaritons and their out-of-equilibrium character allow the observation of different behaviors, at the same time, in different regions of the system. The coexistence of the two behaviors is a peculiar feature typical of cavity polaritons systems distinguishing them from other equilibrium systems like nonlinear optics and atomic condensates.

\section{Conclusions} We have investigated the vortex lattices formation in exciton polariton systems as a function of the polariton density. In the linear regime we have demonstrated the generation of a lattice of vortices in microcavity polariton systems whose size, shape and unit-cell size can be easily controlled and are solely determined by the geometry of the system. When the polariton density is increased, strong polariton-polariton non linear interactions dominate and substantially modify the previously observed array. Our results not only show that the repulsion between polaritons can modify the effect of plane-wave interference and determine the array pattern, but also that the interactions can destroy the topological excitations of opposite winding number by merging one excitation with the other. Our simulations show the correlation between the disappearance of vortex-antivortex pairs and the local onset of the superfluid regime. While our system has a zero total angular momentum and therefore cannot support any single free vortex, Abrikosov-like lattices could be observed by breaking this symmetry therefore opening the way to the observation of mutual vortex-vortex interactions.

\appendix
\section{Sample description}
In our experiment we use a high finesse ($F=3000$) $2\lambda$ GaAs microcavity containing three In${_0.05}$Ga$_{0.95}$As quantum wells. The top/bottom Bragg mirrors are formed by 20 and 24 pairs of alternating layers of GaAs and AlAs with an optical thickness of $\lambda/4$, $\lambda=835$ nm being the wavelength of the confined cavity mode. The Rabi splitting is about 5.1 meV, and the photon lifetime $\tau_C=11$ ps. Since the exciton linewidth is of the order of or slightly lower than the photon linewidth (as considered in the theoretical model), we obtain a polariton lifetime $\tau_{LP}$ of the order of 10-15 ps. This value is consistent with others time-resolved experiments \cite{sanvitto_all-optical_2011}.
 The Bragg mirrors of the cavity form a wedge of a few degrees to allow a fine tuning of the exciton-photon detuning. The microcavity is excited with a continuous-wave single-mode Ti:Sa laser quasi resonant with the lower polariton branch lying in the infrared domain around 837 nm.

\section{Phase extraction procedure}
\begin{figure}
\includegraphics[width=8cm]{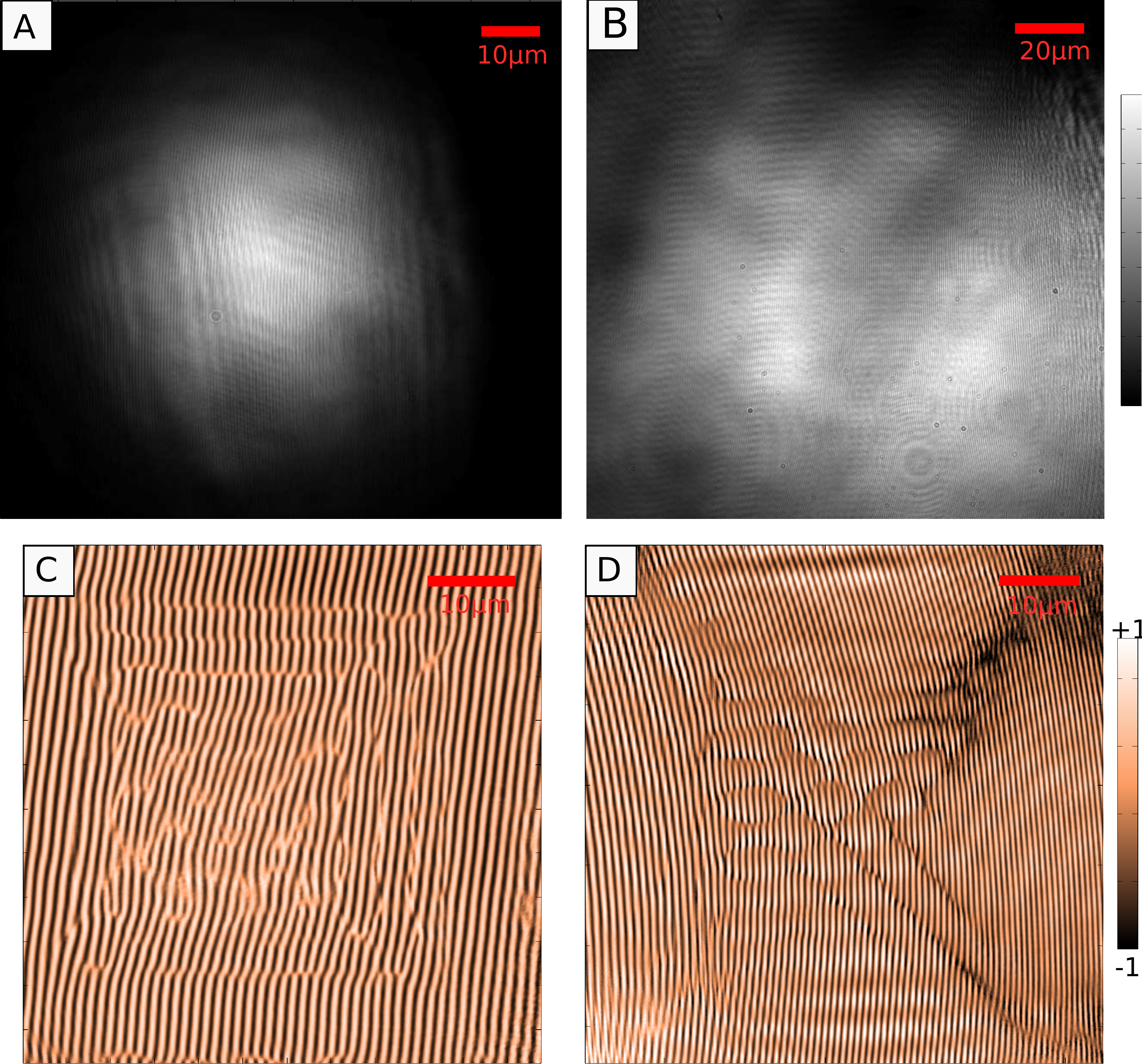}
\caption{\label{fig4s} A-B: Real-space images of typical reference beams in linear greyscale. The reference beam can be either a part of the excitation beam (A) or a small part of the signal greatly enlarged to generate a plane wave (B). The small fringes observable in A are parasitic interferences due to reflection on optical elements of the setup. Due to their very low contrast these fringes do not alter the phase extraction procedure. A and B have been used respectively to extract the phases in the linear and non-linear regimes of figure 2 of the main text. C (D) is the real space interference pattern $\left<\left(E_s+E_r\right)^2\right>$, in linear greyscale, resulting from interferences between A (B) and the signal plotted in figure 2A (2D) of the main text. Since the presence of a vortex is manifested by a fork in the interferogram, one can compare the positions of the forks on C with the vortex distribution represented in fig. 2B of the main text.}
\end{figure}

We detail here the phase extraction procedure. The signal coming from the microcavity $E_s=E_{s0}e^{i\Phi_s}$ (with amplitude $E_{s0}$ and phase $\Phi_s$) is sent into one arm of a Mach-Zehnder interferometer, while a reference beam $E_r=E_{r0}e^{i\phi_r}$ (with amplitude $E_{r0}$ and phase $\Phi_r$) is sent into the other arm of the interferometer. The intensity of the reference beam (figure \ref{fig4s} panels A and B) is adjusted for each set of data in order to maximize the contrast of the interference pattern ($E_{r0}\approx E_{s0}$), and the interference pattern is collected on a CCD camera (figure \ref{fig4s} panels C and D). The detected intensity is therefore proportional to $\left<\left(E_s+E_r\right)^2\right>=E_s^2+E_r^2+2E_{s0}E_{r0}\cos\left(\Phi_s-\Phi_r+\Delta\Phi\right)$ where we assumed a perfect coherence between the two components and where $\Delta\Phi$ is the phase modulation induced by the geometry of the interferometer (angle between arms and lenses). In order to extract the phase diagram we perform a numerical Fourier transform of the interferogram and isolate the off-axis component corresponding to the term $2E_{s0}E_{r0}\cos\left(\Phi_s-\Phi_r+\Delta\Phi\right)$. Then, we perform the inverse Fourier transform from which the phase $\Phi_s-\Phi_r$ (and the amplitude $2E_{s0}E_{r0}$) can be extracted by removing the geometrical term $\Delta\Phi$ that can be measured separately.

\section{Theoretical Model}
A standard way to model the dynamics of resonantly-driven polaritons in a planar microcavity is to use a Gross-Pitaevskii (GP) equation for coupled cavity and exciton fields ($\Psi_C$ and $\Psi_X$) generalized to include the effects of the resonant pumping and decay ($\hbar=1$):

\begin{equation}
\partial_t \begin{pmatrix} \Psi_X \\ \Psi_C \end{pmatrix}
=\begin{pmatrix} 0 \\ F \end{pmatrix} + \left[ H_0 + \begin{pmatrix} g_X|\Psi_X|^2 & 0 \\ 0 & V_C \end{pmatrix}\right] \begin{pmatrix} \Psi_X \\ \Psi_C \end{pmatrix},
\end{equation}

where the single particle polariton Hamiltonian $H_0$ is given by

\begin{equation}
H_0=
\begin{pmatrix}
\omega_X-i\kappa_X & \Omega_R/2 \\ \Omega_R/2 & \omega_C\left(-i\nabla\right)-i\kappa_C \end{pmatrix},
\end{equation}
and
\begin{equation}
\omega_C\left(-i\nabla\right)=\omega_C(0)-\frac{\nabla^2}{2m_c}
\end{equation}
is the cavity dispersion, with the photon mass $m_C= 4\times10^{-5} m_0$ and $m_0$ the bare electron mass. For the simulations we have assumed a flat exciton dispersion relation $\omega_X(k)=\omega_X (0)$. The parameters $\Omega_R$, $\kappa_X$ and $\kappa_C$ are the Rabi frequency and the excitonic and photonic decay rates respectively and have been fixed to values close to experimental ones: $\Omega_R=5.1$ meV, $\kappa_X= 0.04$ meV, and $\kappa_C=0.06$ meV. In this model polaritons are injected into the cavity by a coherent and monochromatic laser field, with pump intensity $f_p$ and a Gaussian spatial profile with width $\sigma$ of 50 $\mu$m : $F(r)=f_p.e^{-r^2/2\sigma^2}$. Here, in contrast to others cases \cite{fraser_2009}, we simulate the polariton system with a simplified two-fields model discarding the role of the excitonic reservoir since our setup is based on the use of a continuous-wave laser with a linewidth orders of magnitudes smaller than the Rabi splitting and therefore the injection of a reservoir of excitons is strongly suppressed. Part of the laser beam profile is set to zero in order to reproduce the effect of the gold-mask. To simulate the effect of not-perfectly reflecting edges of the mask, the pump intensity is set to decay exponentially from the border of the mask toward the centre. The exciton-exciton interaction strength $g_X$ is set to one by rescaling both the cavity and excitonic fields and the pump intensities. Our theoretical results come from the numerical solution of the GP equation over a two-dimensional grid (of $512\times512$ points) in a box with sides of $150\times150\ \mu$m$^2$ using a fifth-order adaptive-step Runge-Kutta algorithm. All the analysed quantities are taken when the system has reached a steady state condition after a transient period of 200 ps.

\begin{acknowledgements}
We would like to thank C. Tejedor for the use of the computational facilities of the Universidad Autonoma de Madrid, L. Martiradonna for the confocal mask, R. Houdr\'e for the sample and Iacopo Carusotto for useful discussions. This work has been partially funded by the Quandyde project of the ANR France, by the POLATOM ESF Research Network Program and by the CLERMONT4 Network Progam. FMM acknowledges financial support from the programs Ramon y Cajal and Intelbiomat (ESF) and MHS from the EPSRC (grant EP/I028900/1). A. B. is member of Institut Universitaire de France (IUF).
\end{acknowledgements}

\bibliography{bibli}{}
\bibliographystyle{ieeetr}
\end{document}